\documentclass[12pt,a4paper]{report}
\usepackage{aims} 
\usepackage[pdftex]{graphicx}
\usepackage{amsmath,amssymb,amsfonts,amsthm,latexsym} 
\DeclareMathOperator\arctanh{arctanh}
\DeclareMathOperator\sech{sech}
\usepackage{moreverb}
\usepackage{subfigure}
\usepackage{braket}
\usepackage{tikz}
\usepackage{pstricks}
\usepackage{pst-plot}
\usepackage[below]{placeins} 
\usepackage[pdftex,colorlinks=true,urlcolor=blue,citecolor=brown,a4paper]{hyperref}

\newcommand {\bea}{\begin{equation}}
\newcommand {\eea}{\end{equation}}

\title{\textbf{ MACROSCOPIC QUANTUM TUNNELING AND COHERENCE OF SPINS}}

\author{ \text{by}\\ Solomon Akaraka Owerre (\textit{solomon.akaraka.owerre@umontreal.ca})\\ 
Universit\`e de Montr\`eal\\\\
{  Supervisor: Professor Manu Paranjape }\\
}
\date{{December 19, 2012}\\{\textit{A dissertation presented to the department of physics in partial fulfilment for predoctoral examination}}\\\vspace{3cm} {\includegraphics[scale=0.2]{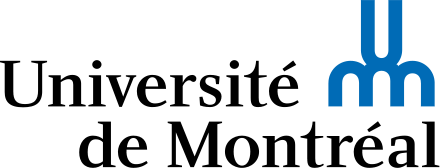}}}  
\begin{document}
\newcommand{\boldnabla}{\mbox{\boldmath$\nabla$}}
\pagestyle{empty}
\maketitle
\pagenumbering{roman}
\tableofcontents
\newpage
%
\pagenumbering{arabic}
\pagestyle{myheadings}


\chapter{Introduction}
\section{Introduction}
 The study of macroscopic quantum tunneling (MQT) and macroscopic quantum coherence (MQC) of spins or magnetization has a long history in physics \cite{D}. The term ``macroscopic" simply means that the system involves very large spin, therefore it can be described using a semi-classical approach. For tunneling to take place, there must be a barrier separating two states. In the case of spins, this mainly involves the tunneling of a macroscopic variable (say $\sigma$ or $M_0= \mu_B \sigma$, where $\mu_B$ is the Bohr magneton) through a barrier between two minima of the effective potential of the system. In MQC,  tunneling between neighbouring degenerate vacua is dominated by the instanton configuration with nonzero topological charge and it leads to an energy level splitting. Tunneling removes the degeneracy of the ground states, and the true ground state is the superposition of the two degenerate ground states. In MQT, tunneling is dominated by the bounce configuration \cite{G} with zero topological charge and it leads to the decay of the metastable states.  The tunneling effect in spin systems occur both in ferromagnetic and anti-ferromagnetic materials \cite{B,C,D}.  In ferromagnetic materials, the macroscopic variables satisfy the well known Landua-Lifshitz differential equation as we shall see soon. 
 
The theoretical problem of tunneling effect involves the calculation of one object--- the tunneling rate (energy splitting). This rate can be calculated semi-classically using two major methods, namely, the WKB method and the instanton method. The instanton method for calculating tunneling amplitude has been studied extensively in one dimension using the imaginary time path integral \cite{G}. For spin systems, however, the imaginary time path integral (coherent-spin-state path integral) gives an additional phase to the transition amplitude. The Euclidean action from this method is first order in time derivative and it has two terms. The first term is the Wess-Zumino term or Berry phase term which is completely imaginary and the second term is the spin (magnetic) anisotropy energy. This term is real and it is the term responsible for the energy barrier between two states. 

The outline of this essay is as follows: In chapter 2, we shall calculate the one instanton contribution to the tunneling rate in small ferromagnetic particles by following closely the method in \cite{D}. We will show that this method uses the incomplete Wess-Zumino term which makes the tunneling rate of half-odd-integer and integer spins to be equivalent. We will further compute the crossover temperature $T_C$, above which the transition process is dominated by thermal hopping over the energy barrier and the transition rate follows the rate $\Gamma = \omega_0\exp\left[-U/k_B T_C\right]$, where $U$ is the energy barrier and $\omega_0$ is the attempt frequency. Below $T_C$, quantum tunneling dominates thermal hopping and one should expect a temperature-independent rate of the form $\Gamma = \omega_0\exp\left[-B\right]$, where $B$ is the Euclidean (imaginary time $t=-i\tau$) action evaluate along the instanton path.

   In chapter 3, we will resolve the problem of the tunneling rate of half-odd-integer and integer spins via coherent-state-path integral method. We will show that the complete Wess-Zumino term leads to a topological phase in the tunneling amplitude. This phase causes a destructive or constructive interference between tunneling paths which leads to the suppression of tunneling rate for half-odd-integer spins (destructive interference) but unsuppressed for integer spins (constructive interference) \cite{A,F}. The suppression of tunneling rate occurs both in ferromagnetic and antiferromagnetic particles. It is as a result of quantum phase interference between tunneling paths of opposite windings. We will also see that if the Hamiltonian is invariant under time-reversal (T) symmetry, the suppression of tunneling rate for half-odd-integer spins is related to the Kramers degeneracy. However, tunneling can also be suppressed with the inclusion of a Zeeman term to the Hamiltonian, in this case the suppression of tunneling is not related to the Kramers degeneracy \cite{K} since this term breaks the T symmetry. 
Finally, in chapter 4, we will make some concluding remarks.

\chapter{Macroscopic quantum tunneling of magnetic moment (spins )}
\section{Tunneling of magnetic moment (spins) in small ferromagnetic particles}

A ferromagnetic material is one in which the elementary magnetic moments or spins spontaneously align below a critical temperature. The magnetic order of ferromagnets generally splits into patterns of magnetic domains in the absence of an external magnetic field. Within a given magnetic domain, the magnetic moments (spins) are all aligned but changes directions at the boundaries between the domains. Thus, each magnetic domain acts like a tiny magnet or grain with large number of magnetic moments but of small volume compared with the size of the magnetic sample.

At equilibrium state, the magnetic domains orient themselves so as to minimize the magnetic anisotropy energy. The general form of the classical energy is given by
\bea
E = C+ \alpha_{ij}M_iM_j + \beta_{ijkl}M_iM_jM_kM_l + \cdots,
\label{2.1}
\eea
where $C$ is a constant, $M_i$ is the magnetic vector, $\alpha_{ij}$ and $\beta_{ijkl}$ are determined by the crystalline anisotropy and by the shape of the magnetic particle. The magnetic vector has at least two or more low-energy directions. Time reversal symmetry gives $\bold{M}\longrightarrow -\bold{M}$, and hence $E(\bold{M}) =E(-\bold{M})$. Therefore, the minimum energy is at least doubly degenerate if it is not at $\bold{M}=0$. Recent investigations have shown that there is a possibility for quantum tunneling of the magnetic vector between these directions which removes the degeneracy of the ground state. In the presence of an external magnetic field, the magnetic domains begin to align with the magnetic field giving rise to a net magnetization, the corresponding energy is given by
\bea
E = C -\bold{M}\cdot \bold{H}+\alpha_{ij}M_iM_j + \beta_{ijkl}M_iM_jM_kM_l + \cdots
\eea
The degeneracy of the energy is thus broken since the magnetic field breaks the time reversal symmetry of the system. 

In this section, we shall calculate the tunneling rate of $\bold{M}$ between degeneracy minima from the classical treatment of the dynamical equations of $\bold{M}$. In the absence of dissipation, the dynamical equation for $\bold{M}$ is given by
\bea
\frac{d \bold{M}}{d t} = \boldsymbol{\tau},\quad \text{where} \quad \boldsymbol{\tau}=-\gamma \bold{M} \times \frac{\delta E}{\delta \bold{M}}, 
\label{2.3}
\eea
often called the Landua-Lifshitz equation. It describes the rotation of a ferromagnetic magnetization in response to torques. The constant $\gamma \equiv ge/2mc$, where $g$ is the gyromagnetic ratio. Expressing $\bold{M}$ in spherical coordinate system i.e $\bold{M} =M_0 \bold{e}_r = M_0\left(\sin\theta \cos\phi, \sin\theta\sin\phi, \cos\theta\right)$,  $\dot{\bold{M}}=\dot{\theta}\mathbf{e}_\theta +\dot{\phi}\sin\theta\mathbf{e}_\phi$, $\boldsymbol{\tau}=-\gamma\mathbf{e}_\phi\partial E/\partial\theta + \gamma\mathbf{e}_\theta \partial E/\sin\theta\partial\phi$, one can then obtain Eq. \eqref{2.3} directly from the Minkowski action
\bea
S_M = \int dt \left[(M_0/\gamma)\dot{\phi}\cos\theta -E(\theta,\phi)\right].
\label{2.4}
\eea
Introducing the canonical variables
\bea
x=\phi, \quad p = (M_0/\gamma)\cos\theta= hS_z,
\label{2.5}
\eea
where $S_z$ is the $z$ projection of the total spin of the particle, the Lagrangian of the system can be written as
\bea
L = p\dot{x}-E.
\label{2.6}
\eea
The semi-classical tunneling problems are often treated by switching to the imaginary time $(t=-i\tau)$ action or Euclidean action. The corresponding Euclidean action of \eqref{2.4} is
\bea
S_E=-iS_M = \int d\tau \left[-i(M_0/\gamma)\dot{\phi}\cos\theta +E(\theta,\phi)\right].
\label{2.7}
\eea
Minimizing the action \eqref{2.7} with respect to $\theta$ and $\phi$, we have
\begin{align}
  i(M_0/\gamma)\dot{\bar{\theta}}\sin\bar{\theta} &= \frac{\partial E}{\partial \bar{\phi}},\label{2.8}\\
  i(M_0/\gamma)\dot{\bar{\phi}}\sin\bar{\theta} &= -\frac{\partial E}{\partial \bar{\theta}},
\label{2.9}
\end{align}
where $\bar{\theta}$ and $\bar{\phi}$ are the classical paths that minimize the action. Since the anisotropy energy $E$ is real, these two equations are inconsistent unless either $\bar{\theta}$ or $\bar{\phi}$ is imaginary.
\begin{figure}[h!]
\centering
 \includegraphics[width=2.5in]{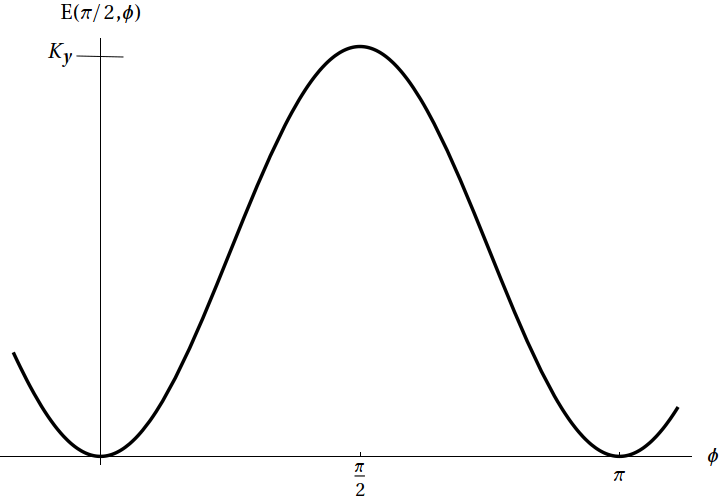}
\caption{Anisotropy energy vs $\phi$ at $\theta = \pi/2$.} \label{fig2.1}
\end{figure}
\subsection{Models for small ferromagnetic particles}
We shall start with the analysis of a small ferromagnetic particle with $XOY$-easy-plane anisotropy with easy axis along the $x$-direction in the plane, medium axis along the $y$-direction and hard axis along the $z$-direction considered as model I in \cite{D}. The classical anisotropy energy $E$ is of the form
\bea
E(\hat{\bold{n}})=E(\theta,\phi)= K_zM_z^2 +K_yM_y^2 = K_zM_0^2\cos^2\theta + K_yM_0^2\sin^2\theta\sin^2\phi,
\label{2.10}
\eea
where $\hat{\bold{n}}$ is the magnetization direction and $K_z>K_y>0$ are the anisotropy constants. The ground state of the system corresponds to $\bold{M}$ pointing in one of the two directions parallel to the $x$-axis with $\theta = \pi/2$, $\phi =0, \pi$ as shown in fig.\eqref{fig2.1}. 
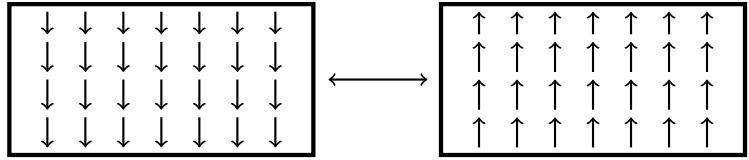
\begin{figure}[h!]
\centering
\mbox{\subfigure{\begin{tikzpicture}
\draw[ultra thick](0,0) rectangle (4,2);
\draw[<-,thick](3.5,0.1)--(3.5,0.5);
\draw[<-,thick](3,0.1)--(3,0.5);
\draw[<-,thick](2.5,0.1)--(2.5,0.5);
\draw[<-,thick](2,0.1)--(2,0.5);
\draw[<-,thick](1.5,0.1)--(1.5,0.5);
\draw[<-,thick](1,0.1)--(1,0.5);
\draw[<-,thick](0.5,0.1)--(0.5,0.5);
\draw[<-,thick](3.5,0.6)--(3.5,1.0);
\draw[<-,thick](3.0,0.6)--(3.0,1.0);
\draw[<-,thick](2.5,0.6)--(2.5,1.0);
\draw[<-,thick](2,0.6)--(2,1.0);
\draw[<-,thick](1.5,0.6)--(1.5,1.0);
\draw[<-,thick](1,0.6)--(1,1.0);
\draw[<-,thick](0.5,0.6)--(0.5,1.0);
\draw[<-,thick](3.5,1.1)--(3.5,1.5);
\draw[<-,thick](3.0,1.1)--(3.0,1.5);
\draw[<-,thick](2.5,1.1)--(2.5,1.5);
\draw[<-,thick](2,1.1)--(2,1.5);
\draw[<-,thick](1.5,1.1)--(1.5,1.5);
\draw[<-,thick](1,1.1)--(1,1.5);
\draw[<-,thick](0.5,1.1)--(0.5,1.5);
\draw[<-,thick](3.5,1.6)--(3.5,1.9);
\draw[<-,thick](3.0,1.6)--(3.0,1.9);
\draw[<-,thick](2.5,1.6)--(2.5,1.9);
\draw[<-,thick](2,1.6)--(2,1.9);
\draw[<-,thick](1.5,1.6)--(1.5,1.9);
\draw[<-,thick](1,1.6)--(1,1.9);
\draw[<-,thick](0.5,1.6)--(0.5,1.9);
\draw[<->,thick] (4.2,1)--(5.5,1);
\end{tikzpicture}}
\subfigure{\begin{tikzpicture}
\draw[ultra thick](0,0) rectangle (4,2);
\draw[->,thick](3.5,0.1)--(3.5,0.5);
\draw[->,thick](3,0.1)--(3,0.5);
\draw[->,thick](2.5,0.1)--(2.5,0.5);
\draw[->,thick](2,0.1)--(2,0.5);
\draw[->,thick](1.5,0.1)--(1.5,0.5);
\draw[->,thick](1,0.1)--(1,0.5);
\draw[->,thick](0.5,0.1)--(0.5,0.5);
\draw[->,thick](3.5,0.6)--(3.5,1.0);
\draw[->,thick](3.0,0.6)--(3.0,1.0);
\draw[->,thick](2.5,0.6)--(2.5,1.0);
\draw[->,thick](2,0.6)--(2,1.0);
\draw[->,thick](1.5,0.6)--(1.5,1.0);
\draw[->,thick](1,0.6)--(1,1.0);
\draw[->,thick](0.5,0.6)--(0.5,1.0);
\draw[->,thick](3.5,1.1)--(3.5,1.5);
\draw[->,thick](3.0,1.1)--(3.0,1.5);
\draw[->,thick](2.5,1.1)--(2.5,1.5);
\draw[->,thick](2,1.1)--(2,1.5);
\draw[->,thick](1.5,1.1)--(1.5,1.5);
\draw[->,thick](1,1.1)--(1,1.5);
\draw[->,thick](0.5,1.1)--(0.5,1.5);
\draw[->,thick](3.5,1.6)--(3.5,1.9);
\draw[->,thick](3.0,1.6)--(3.0,1.9);
\draw[->,thick](2.5,1.6)--(2.5,1.9);
\draw[->,thick](2,1.6)--(2,1.9);
\draw[->,thick](1.5,1.6)--(1.5,1.9);
\draw[->,thick](1,1.6)--(1,1.9);
\draw[->,thick](0.5,1.6)--(0.5,1.9);
\end{tikzpicture} }}
\caption{Degenerate energy minima of a single-domain ferromagnet grain; the spin tunnels between these configurations.} \label{fig2.2}
\end{figure}
 
 Before we proceed further, let us point out that not all quantum spin Hamiltonian possess quantum tunneling. As an example let us consider the simplest anisotropy energy  
\begin{equation}
E = K_z S_z^2 -\gamma H S_z =K_zM_0^2\cos^2\theta -\gamma H M_0 \cos\theta
\label{2.11}
\end{equation}
where $K_z>0$ is an anisotropy constant, $H$ is the magnetic field and $\gamma$ is related to the $g$ factor. Eq. \eqref{2.11} corresponds to the quantum spin Hamiltonian
\begin{equation}
\hat{H} = K_z \hat{S_z}^2 -\gamma H \hat{S_z}  
\label{2.12}
\end{equation}
 A quick glance at \eqref{2.12} shows that the Hamiltonian commutes with $\hat{S_z}  $. Therefore $\hat{S_z}$ is a conserved quantum number and the above Hamiltonian cannot possess any quantum transition. This is easily seen by lack of instanton solution of \eqref{2.8} and \eqref{2.9} using \eqref{2.11}. Thus, the minimal model that possesses quantum tunneling requires terms in the Hamiltonian that do not commute with it.
 Returning to Eq.\eqref{2.10}, in order to compute the tunneling rate via instanton method, we first find the solution of the classical equations of motion \eqref{2.8} and \eqref{2.9}. One can easily derive the  conservation of energy directly from these two equations by multiplying \eqref{2.8} by $\dot{\bar{\phi}}$ and \eqref{2.9} by $\dot{\bar{\theta}}$ and subtracting the resulting equations:
\begin{equation}
\frac{dE}{d\tau}=\dot{\bar{\phi}}\frac{\partial E}{\partial \bar{\phi}} + \dot{\bar{\theta}}\frac{\partial E}{\partial \bar{\theta}}=0 \Longrightarrow E =0.
\label{2.13}
\end{equation}
The energy remains zero along the instanton trajectory. Using \eqref{2.10} and \eqref{2.13} we obtain an expression for $\cos\bar{\theta}$ in terms of $\bar{\phi}$:
\begin{equation}
\cos\bar{\theta} =\frac{i\lambda^{1/2}\sin\bar{\phi}}{\sqrt{1-\lambda\sin^2 \bar{\phi}}},
\label{2.14}
\end{equation}
where $\lambda= \ K_y/K_z$.
Substituting \eqref{2.14} into \eqref{2.9} we obtain an equation for $\bar{\phi}$ only:
\begin{equation}
\dot{\bar{\phi}}^2 =\omega_0^2\sin^2 \bar{\phi}(1-\lambda\sin^2 \bar{\phi}),
\label{2.15}
\end{equation}
where $\omega_0 = (2M_0/\gamma)(K_zK_y)^{1/2}$. Integrating \eqref{2.15} we obtain the instanton solution
\begin{equation}
\bar{\phi}(\tau) =\pm\arccos\left(\frac{(\sqrt{1-\lambda})\tanh(\omega_0\tau)}{\sqrt{1-\lambda\tanh^2(\omega_0\tau)}}\right).
\label{2.16}
\end{equation}
Notice that the instanton  is a real function of $\tau$ which corresponds to the switching of $\bold{M}$ from $\bar{\phi} = 0$ at $\tau=-\infty$ to $\bar{\phi} = \pi$ at $\tau=\infty$. The action for this path can be obtained by substituting equations \eqref{2.13} and \eqref{2.14} into \eqref{2.7}, this gives
\begin{equation}
B=(M_0/\gamma)\sqrt{\lambda} \int_{0}^{\pi}d\bar{\phi}\thinspace\frac{\sin\bar{\phi}}{\sqrt{1-\lambda\sin^2\bar{\phi}}} = \ln\left(\frac{1+\sqrt{\lambda}}{1-\sqrt{\lambda}}\right)^{M_0/\gamma}. 
\label{2.17}
\end{equation}
The one instanton contribution to the tunneling rate is given by the expression \cite{D,I,G}
\bea
\mathcal{P} \propto \exp\left(-B/\hbar\right)= \left(\frac{1-\sqrt{\lambda}}{1+\sqrt{\lambda}}\right)^{M_0/\hbar\gamma}.
\label{2.18}
\eea
Consider the limit $K_z\rightarrow K_y\rightarrow K$, in this limit $E\rightarrow K(M_z^2 +M_y^2)= K\bold{M}^2-KM_x^2$ which clearly commutes with $M_x$. Thus we expect the tunneling rate to go to zero. This is obviously the case since $\lambda\rightarrow 1$ in this limit, therefore $\mathcal{P}\rightarrow 0$. Recent experiments suggest that this tunneling effect is observable in particles with several thousand of large spins i.e $M_0/\hbar\gamma$ is very large, so the tunneling rate \eqref{2.18} can be observed when $\lambda <<1$ which implies that $K_z>>K_y$.

Consider the classical anisotropy energy
\begin{equation}
E(\hat{\bold{n}})=E(\theta,\phi)=-K_zM_z^2 + K_y M_y^2 -\gamma\bold{M}\cdot\bold{H},
\end{equation}
where $\bold{H}$ is applied along the $z$-axis. This model is considered as model III in \cite{D}.
Up to a constant, this is equivalent to
\begin{equation}
E(\theta,\phi)=M_0^2( K_z  + K_y \sin^2\phi)\sin^2\theta -  \gamma M_0H(1-\cos\theta).
\label{2.20}
\end{equation}
 It can be easily shown that for $\phi = 0$ , and provided $ H<H_c = 2K_z M_0/\gamma$, the energy has two local minima at $\theta =0$ and $\theta =\pi$ and a maximum at $\cos\theta_1 = H/H_c$. The energy barrier between the minima is $E(\theta_1,0)= U = M_0^2K_z\epsilon^2$, where $\epsilon= 1-H/H_c$. All our calculation for this model will be done to leading order in $\epsilon$. In the limit $\epsilon \rightarrow 0$, both $\theta$ and $\phi$ are small, we get from \eqref{2.20}
\begin{equation}
E(\theta,\phi) =M_0^2K_z\left(\epsilon\theta^2 -\frac{\theta^4}{4}\right)  + M_0^2K_y \phi^2\theta^2 +\cdots
\label{2.21}
\end{equation}
\begin{figure}[h!]
\centering
 \includegraphics[width=3in]{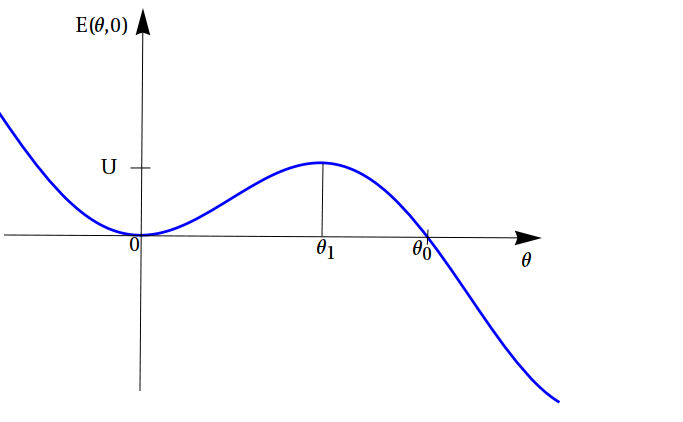} 
\caption{The anisotropy energy $E(\theta, \phi=0)$ with a metastable state at $\theta=0$. Here $\theta_0=2\sqrt{\epsilon}$, $\theta_1=\sqrt{2\epsilon}$ and $U=K_zM_0^2\epsilon^2$.} \label{fig2.3}
\end{figure}
In order to find the bounce solution, we use the conservation of energy \eqref{2.13} to express $\bar{\phi}$ in terms of $\bar{\theta}$:
\begin{equation}
 \bar{\phi}  = i \left[\frac{K_z}{K_y}(\epsilon-\bar{\theta}^2/4)\right]^{1/2}.
\label{2.22}
\end{equation}
We can now eliminate $\bar{\phi}$ from \eqref{2.8} using \eqref{2.22} and the resulting equation is
\begin{equation}
 \dot{\bar{\theta}}^2  = \omega_0^2(\epsilon \bar{\theta}^2-\bar{\theta}^4/4) ,
\label{2.23}
\end{equation}
where $\omega_0 = (2M_0/\gamma)(K_zK_y)^{1/2}$. Integrating we obtain the bounce solution
\begin{equation}
  \bar{\theta}(\tau)  = \theta_0\sech(\omega_0\sqrt{\epsilon}\tau), 
\label{2.24}
\end{equation}
which corresponds to interpolation of $\bar{\theta}$ from $\bar{\theta}=0$ at $\tau=-\infty$ to $\bar{\theta}=\theta_0 =2\sqrt{\epsilon}$ at $\tau =0$, and then back to $\bar{\theta}=0$ at $\tau =\infty$. 
The action for the bounce path is given by
\begin{equation}
\begin{split}
B &=-i(M_0/\gamma)\int_{-\infty}^{\infty}d \tau \dot{\bar{\phi}}\cos\bar{\theta}\approx \frac{iM_0}{2\gamma}\int_{-\infty}^{\infty}d \tau \dot{\bar{\phi}}\bar{\theta}^2 \\
& =\frac{M_0}{4\gamma}\left(\frac{K_z}{K_y}\right)^{1/2}\int_{0}^{\theta_0}\frac{\bar{\theta}^3} {\sqrt{\epsilon - \bar{\theta}^2/{4}}}\thinspace d\bar{\theta} =   8M_0/3\hbar\gamma\left( K_z/K_y\right)^{1/2}\epsilon^{3/2}.
\end{split}
\label{2.25} 
\end{equation}
The tunneling rate in this case is given by the expression 
\bea
\mathcal{P} \propto \exp\left(-B/\hbar\right)= \exp\left[-8M_0/3\hbar\gamma\left( K_z/K_y\right)^{1/2}\epsilon^{3/2}\right].
\label{2.26}
\eea
At high temperature $T>T_c$, where $T_c$ is the crossover temperature, quantum transition is dominated by thermal hopping over the barrier. Thus the transition rate follows the law:
\begin{equation}
\mathcal{P} \propto e^{-U/K_BT_c} .
\label{2.27}
\end{equation}
where $K_B$ is the Boltzmann constant and $U=M_0^2K_z\epsilon^2$ is the height of the barrier. Comparing \eqref{2.26} and \eqref{2.27} we obtain the crossover temperature
\begin{equation}
  K_BT_c= \hbar U/B  = 3M_0\hbar\gamma(K_zK_y)^{1/2}\sqrt{\epsilon}/8.
\label{2.28}
\end{equation}
\section{Tunneling of spins in small antiferromagnetic particles}

The quantum mechanical coupling between magnetic moments in some materials is such that adjacent magnetic moments tend to line up along opposite directions. The long-range order in these materials  can be described in terms of two opposing ferromagnetic sublattices which is the simplest form of the Ne\'el model. If the net magnetizations of the two sublattices are equal, the material is called an antiferromagnet. In the absence of an external magnetic field, the magnetization of the two sublattices are opposite to each other i.e $\bold{M}_1= -\bold{M}_2$, so the total magnetization cancels, yielding no net magnet moment. Antiferromagnetic order is characterized by the Ne\'el vector of unit length
\bea
\bold{N}= \frac{\bold{M}_1-\bold{M}_2}{2M_0}.
\label{2.29}
\eea
In this case, we are interested in the quantum tunneling of $\bold{N}$ between two opposite orientations, $\ket{\uparrow \downarrow}$ and $\ket{\downarrow\uparrow}$. In the absence of an external magnetic field, the Lagrangian of the uniaxial antiferromagnet is \cite{R}

\bea
L = \int d^3 x\left[\frac{\chi_{\perp}}{2\gamma^2}\left(\frac{d\bold{N}}{dt}\right)^2-\frac{\alpha}{2}\left(\frac{\partial N_i}{\partial x_i}\right)^2 +\frac{1}{2}K\left(\bold{n}\cdot\bold{N}\right)^2\right],
\label{2.30}
\eea
where $\chi_{\perp}$ is the perpendicular susceptibility with respect to the equilibrium orientation of $\bold{N}$ along the anisotropy axis $\bold{n}$, $\gamma= e/mc$, $\alpha$ and $K$ are the exchange interaction constants correspondingly. Now for a small particle, the spatial derivatives of $\bold{N}$ are suppressed by the exchange interaction, so $\bold{N}$ may depend only on time. Representing $\bold{N}$ in spherical coordinate system $(\bold{n}\cdot\bold{N}=\cos\theta)$, we obtain from \eqref{2.30}
\bea
L = V\left\lbrace\frac{\chi_{\perp}}{2\gamma^2}\left[\left(\frac{d\theta}{dt}\right)^2 + \left(\frac{d\phi}{dt}\right)^2\sin^2 \theta \right] -\frac{1}{2}K\sin^2\theta\right\rbrace,
\label{2.31}
\eea
where $V$ is the volume of the particle. We have added a constant term in \eqref{2.31} for convenience.

The degenerate classical minimum energy $E=0$ corresponds to the equilibrium orientations of $\bold{N}$ at $\theta =0$ and $\theta =\pi$. The tunneling rate between these two degenerate minima can be found by switching to the imaginary time version of \eqref{2.31}:
\bea
S_E = V\int d\tau \left\lbrace\frac{\chi_{\perp}}{2\gamma^2}\left[\left(\frac{d\theta}{d\tau}\right)^2 + \left(\frac{d\phi}{d\tau}\right)^2\sin^2 \theta \right]+\frac{1}{2}K\sin^2\theta\right\rbrace,
\label{2.32}
\eea
The equations of motion from the Euclidean action are
\begin{align}
\frac{d\bar{\phi}}{d\tau}\sin^2\bar{\theta}&=\text{const},\\
\frac{\chi_{\perp}}{\gamma^2}\frac{d^2\bar{\theta}}{d\tau^2}&= \left(K +\frac{\chi_{\perp}}{\gamma^2}\left(\frac{d\bar{\phi}}{d\tau}\right)^2\right)\sin\bar{\theta}\cos\bar{\theta}.
\end{align}

A classical rotation of $\bold{N}$ may occur in any plane $\phi =$ const. Thus, we have
\bea
2\frac{d^2\bar{\theta}}{d\tau^2}= \omega_0^2\sin 2\bar{\theta},
\eea
where $\omega_0 =\gamma\left(K/\chi_\perp\right)^{1/2}$. Integrating once we obtain  
\bea
\frac{\chi_{\perp}}{2\gamma^2}\left(\frac{d\bar{\theta}}{d\tau}\right)^2-\frac{1}{2}K\sin^2\bar{\theta} = E=0,
\eea
and the corresponding instanton solution is
\bea
\bar{\theta}(\tau) =2\arctan\left[\exp(\omega_0\tau)\right].
\eea
This solution corresponds to a subbarrier rotation of $\bold{N}$ from $\theta =0$ at $\tau=-\infty$ to $\theta=\pi$ at $\tau =\infty$. The action for this path is easily obtained from \eqref{2.32} as:
\bea
B = V\int d\tau \left\lbrace\frac{\chi_{\perp}}{2\gamma^2} \left(\frac{d\bar{\theta}}{d\tau}\right)^2  +\frac{1}{2}K\sin^2\bar{\theta}\right\rbrace = KV\int d\tau  \sin^2\bar{\theta} =2V\frac{\sqrt{\chi_{\perp}K}}{\gamma}.
\label{2.35}
\eea
Thus, the tunneling rate is
\bea
\mathcal{P}\propto \exp(-B/\hbar) =\exp(-2V\sqrt{\chi_{\perp}K}/\gamma\hbar).
\eea
At high temperature the tunneling rate is dominated by the thermal hopping over the energy barrier $U = \frac{1}{2}KV$ and the critical (crossover) temperature  is given by
\bea
T_c =\frac{\gamma\hbar}{2K_B}\sqrt{\frac{K}{\chi_{\perp}}}.
\eea
\chapter{Spin-parity effect in macroscopic quantum tunneling of spin systems}
\section{Suppression of tunneling in half-odd-integer-spin   ferromagnetic particles }
We carelessly omitted the complete topological phase term (that is the Wess-Zumino phase or Berry phase) in the previous chapter. This phase is responsible for the suppression of tunneling in spin systems. In this section, we will show how the complete phase comes from setting up a spin-coherent-state path integral. In the spin-coherent-state formalism, we will see that in the absence of a magnetic field, the quantum tunneling of magnetization direction is spin-parity dependent. It is completely suppressed if the total spin of the magnetic particle is half integral (fermions) but is allowed in integral-spin (bosons) particles. The  quenching of tunneling rate in the absence of a magnetic field is related to Kramers theorem which states that if the Hamiltonian of a system possesses time reversal symmetry, then the ground state energy is at least doubly degenerate. We shall show that the quenching of the tunneling amplitude has a topological origin, the topological phase can lead to destructive quantum interference between different tunneling paths and hence leads to the vanishing of the tunneling amplitude. Moreover, quenching of tunneling still persists in the presence of a magnetic field at certain value of the field, in this case the suppression of tunneling is not related to the Kramers theorem since the magnetic field breaks the time reversal symmetry of the problem.
\subsection{Spin-coherent-state path integral formalism}
Consider a single spin  particle $s$. Let us define the Hilbert space of $SU(2)$ as: 
\begin{equation*}
\left\lbrace\ket{s, m}, m = -s,-s+1, \cdots,s-1, s
; s = \text{integer or half-odd-integer}\right\rbrace.
\end{equation*}
There are $2s +1$ states and $\ket{s,m}$ is a simultaneous eigenstate of the $SU(2)$ Casimir operators $\bold{\hat{S}}^2$ and $\hat{S_z}$:
\begin{equation}
\begin{split}
\bold{\hat{S}}^2\ket{s,m} &= s(s+1)\ket{s,m},\\
\hat{S_z}\ket{s,m} &= m\ket{s,m}.
\label{5.0}
\end{split}
\end{equation}
Similar to the case of harmonic oscillator, we can obtain the state $\ket{s,m}$ by applying the operator $\hat{S}_- = \hat{S}_x-i\hat{S}_y$,  $p$ times to the state  with maximum value of $m$,  i.e $\ket{s,s} \equiv \ket{0}$:
\begin{equation}
\begin{split}
\left(\hat{S}_-\right)^{p}\ket{0}={{2s}\choose{p}}^{\frac{1}{2}}{p!}\ket{p},
\label{5.1}
\end{split}
\end{equation}
where $\ket{p}$ is such that
\bea
\hat{S}_z \ket{p}= \left(s-p\right)\ket{p}.
\eea 
In order to be consistent with \eqref{5.0}, we can identify  the state $\ket{p}$ and the eigenvalue $p$ as $\ket{p}= \ket{s,m}$ and $p=s-m$ with $0\leq p \leq 2s$. One can verify that Eq.\eqref{5.1} gives the correct expression for $p=1 \thinspace(\text{i.e},\thinspace m=s-1)$ by comparing it with the well known relation 
\bea
\hat{S}_- \ket{s,m}=\sqrt{(s+m)(s-m+1)}\ket{s,m-1}
\eea
for $m=s$.

Consider the state
\begin{equation}
\ket{\mu} \equiv N^{-1/2}\exp(\mu \hat{S}_-)\ket{0} = N^{-1/2}\sum_{p=0}^{2s}{{2s}\choose{p}}^{\frac{1}{2}}\mu^{p}\ket{p},
\label{5.2}
\end{equation}
where $\mu$ runs over the complex plane and $N$ is a normalization factor. The normalization factor can be obtained from the condition:
\begin{equation}
\braket{\mu|\mu}  = N^{-1}\sum_{p=0}^{2s}{{2s}\choose{p}}\lvert\mu\rvert^{2p} =N^{-1}\left(1+\lvert\mu\rvert^2\right)^{2s}=1.
\label{5.3}
\end{equation}
Hence, the normalized state is
\bea
\ket{\mu} = \left(1+\lvert\mu\rvert^2\right)^{-s}\exp(\mu \hat{S}_-)\ket{0}.
\label{5.4}
\eea
The overlap between two states $\ket{\mu^{\prime}}$ and $\ket{\mu}$ is
\bea
\braket{\mu^{\prime}|\mu}=\frac{\left(1+\mu^{\prime}\mu\right)^{2s}}{\left(1+\lvert\mu^{\prime}\rvert^2\right)^{s}\left(1+\lvert\mu\rvert^2\right)^{s}},
 \label{5.5}
 \eea
and the completeness relation is
\begin{equation}
 \frac{2s+1}{\pi}\int\frac{d^2\mu}{(1+\lvert\mu\rvert^2)^2}\ket{\mu}\bra{\mu}= \sum_{p=0}^{2s}\ket{p}\bra{p} =1.
\label{5.6}
\end{equation}
In terms of the spherical parametrization $\theta$ and $\phi$,  \thinspace $0\leq \theta < \pi$, \thinspace $0\leq \phi < 2\pi$,  we have \cite{T,M}
\bea
\mu =\tan\left(\frac{1}{2}\theta\right)e^{i\phi},
\eea
 where $\theta$ and $\phi$ correspond to the points on the sphere that are stereographically projected to $\mu$. Then, Eq.\eqref{5.4}--Eq.\eqref{5.6} can be written as
\begin{equation}
\ket{\mu}\equiv \ket{\theta, \phi} \equiv \ket{\bold{\hat{n}}} = \left(\cos\frac{1}{2}\theta\right)^{2s}\exp\left\lbrace\tan\left(\frac{1}{2}\theta\right)e^{i\phi}\hat{S}_-\right\rbrace\ket{0}.
\label{5.7}
\end{equation}
The overlap becomes
\begin{equation}
 \braket{\bold{\hat{n}}^{\prime}|\bold{\hat{n}}}=\left\lbrace\cos\frac{1}{2}\theta\cos\frac{1}{2}\theta^{\prime} +\sin\frac{1}{2}\theta\sin\frac{1}{2}\theta^{\prime}e^{i(\phi-\phi^{\prime})}\right\rbrace^{2s},
\label{5.8}
\end{equation}
it follows that, after a lot of algebra
\begin{equation}
 \lvert\braket{\bold{\hat{n}}^{\prime}|\bold{\hat{n}}}\rvert =\left(\frac{1}{2}(1+\bold{\hat{n}}^{\prime}\cdot\bold{\hat{n}}) \right)^{2s}.
\label{5.9}
\end{equation}
For infinitesimal separated angles, $\delta \phi = \phi^{\prime}-\phi$, $\delta \theta = \theta^{\prime} -\theta$, the overlap \eqref{5.8} becomes
\begin{equation}
 \braket{\bold{\hat{n}}^{\prime}|\bold{\hat{n}}}=1 -i s\delta\phi(1-\cos\theta),
\label{6.1}
\end{equation}
and for large $s$
\begin{equation}
\braket{\bold{\hat{n}}^{\prime}|\bold{\hat{S}}|\bold{\hat{n}}}  =s\left[\bold{\hat{n}} +O\left(\sqrt{s} \right)\right]\braket{\bold{\hat{n}}^{\prime}|\bold{\hat{n}}} 
\label{5.9a},
\end{equation}
where $\bold{\hat{n}}= (\sin\theta\cos\phi, \sin\theta\sin\phi, \cos\theta)$.
The completeness relation becomes
\begin{equation}
 \frac{2s+1}{4\pi}\int d\theta \thinspace d\phi \thinspace\sin \theta \ket{\bold{\hat{n}}}\bra{\bold{\hat{n}}} = \frac{2s+1}{4\pi}\int d\bold{\hat{n}} \ket{\bold{\hat{n}}}\bra{\bold{\hat{n}}}=1.
\label{6.0}
\end{equation}

Having developed all these tools, let us now construct the path integral representation for the transition amplitude between two spin configurations. Following the usual procedure \cite{H}, we discretize the time interval into $N$ identical pieces of length $\epsilon =\beta/N$ and insert a complete set of states at each site,
\begin{equation}
 \braket{\bold{\hat{n}}_b|e^{- \beta \hat{H}(\bold{\hat{S}})}|\bold{\hat{n}}_a}=\braket{\bold{\hat{n}}_b|\left(e^{- \epsilon \hat{H}(\bold{\hat{S}})}\right)^{N}|\bold{\hat{n}}_a} =\left(\prod_{i=1}^{N-1}\int  \frac{2s+1}{4\pi} d \hat{n}_i\prod_{j=0}^{N-1}\braket{\bold{\hat{n}}(\tau_{j+1})| e^{- \epsilon \hat{H}(\bold{\hat{S}})}|\bold{\hat{n}}(\tau_j)}\right),
\label{6.2}
\end{equation}
where $\tau_j = \tau +j\epsilon$, $\ket{\bold{\hat{n}}(\tau_0)} = \ket{\bold{\hat{n}}_a}$ and $\ket{\bold{\hat{n}}(\tau_N)} = \ket{\bold{\hat{n}}_b}$. In the limit of large $s$, we use \eqref{6.1} and \eqref{5.9a} and write the right hand side of \eqref{6.2} as
\begin{equation}
\begin{split}
\prod_{j=0}^{N-1}\braket{\bold{\hat{n}}(\tau_{j+1})|e^{- \epsilon \hat{H}(\bold{\hat{S}})}|\bold{\hat{n}}(\tau_j)} &=\prod_{j=0}^{N-1} \left(1-\epsilon \hat{H}(s\bold{\hat{n}}(\tau_j))\right) \braket{\bold{\hat{n}}(\tau_{j+1})|\bold{\hat{n}}(\tau_j)} + O(\epsilon^2) \\
&= \prod_{j=0}^{N-1} \left(1-\epsilon \hat{H}(s\bold{\hat{n}}(\tau_j))\right) \left[1-is\delta\phi(\tau_j)(1-\cos\theta(\tau_j))\right]+ O(\epsilon^2)\\&= \exp\left[-\epsilon\sum_{j=0}^{N-1}\left\lbrace is \frac{\phi(\tau_{j+1})-\phi(\tau_j)}{\epsilon}(1-\cos\theta(\tau_j)) + H(s\bold{\hat{n}}(\tau_j))\right\rbrace\right]
.
\label{6.3}
\end{split}
\end{equation}
 In the continuum limit $N\longrightarrow \infty$, $\epsilon \longrightarrow 0$ we have
\begin{equation}
 \braket{\bold{\hat{n}}_b|e^{- \beta \hat{H}(\bold{\hat{S}})}|\bold{\hat{n}}_a}=   \int \mathcal{D}\bold{\hat{n}}(\tau)e^{-S_E},
\label{6.4}
\end{equation}
where
\bea
\mathcal{D}\bold{\hat{n}}(\tau) = \mathcal{N}\prod_{i=1}^{N-1}d \hat{n}(\tau_i),
\eea
\begin{equation}
 S_E = S_{WZ} + \int_{0}^{\beta}d\tau H\left[s\bold{\hat{n}}(\tau)\right]=\int_{0}^{\beta} d \tau \mathcal{L}_E,
\label{6.5},
\end{equation}
where the  Euclidean Lagrangian is
\bea
\mathcal{L}_E = is\dot{\phi}(1-\cos\theta ) + E\left(\theta, \phi \right).
\label{6.8}
\eea
The coordinates $\left(\theta, \phi \right)$ label the coherent spin state $\ket{\theta, \phi }$ for a particle with spin $s$. It is related to the direction of the unit vector $\bold{\hat{n}}$ on a two-sphere. The first term in Eq.\eqref{6.8} is the full Wess-Zumino term which takes into account the fact that the original quantum spin satisfies the algebra of the rotation group. The semi-classical energy $E$ is the expectation value $\braket{\theta,\phi|\hat{H}|\theta,\phi}$ of the Hamiltonian operator $\hat{H}$. We will be interested in the case where $\ket{\bold{\hat{n}}_a}$ and $\ket{\bold{\hat{n}}_b}$ are classical degenerate ground states separated by an energy barrier such that $ \braket{\bold{\hat{n}}_a|\hat{H}|\bold{\hat{n}}_a}$ and $ \braket{\bold{\hat{n}}_b|\hat{H}|\bold{\hat{n}}_b}$ are the smallest possible expectation values of $\hat{H}$.
\subsection{Ferromagnetic models using the full Wess-Zumino term}
We shall begin the analysis in this formalism by re-examining the tunneling behaviour considered as ``model I" in the previous chapter. The classical anisotropy energy corresponds to the quantum spin Hamiltonian
\bea
\hat{H} = K_z\hat{S_z^2} +K_y\hat{S_y^2}.
\label{3.1}
\eea
We want to compute the tunneling of the magnetization direction $\bold{\hat{n}}$ between its two equivalent directions corresponding to the coherent states $\ket{\bold{\hat{n}}_a}=\ket{\theta = \pi/2,\thinspace \phi =0}$ and $\ket{\bold{\hat{n}}_b}=\ket{\theta = \pi/2,\thinspace \phi =\pi}$ . Using the spin coherent state path integral developed above, the transition amplitude is 
\begin{equation}
 \braket{\bold{\hat{n}}_b|e^{- \beta \hat{H}/\hbar }|\bold{\hat{n}}_a}=\braket{\phi=\pi|e^{- \beta \hat{H}/\hbar }|\phi=0}=   \int  \mathcal{D}\phi(\tau)\mathcal{D}\cos\theta(\tau) e^{-S_E/\hbar},
\label{7.1}
\end{equation}
and the Euclidean action is  
\begin{equation}
S_E = \int_{-\beta/2}^{\beta/2}d\tau\thinspace \left[is\dot{\phi}(1-\cos\theta )+ E\left(\theta, \phi \right)\right], 
\label{7.1b}
\end{equation}
The classical anisotropy energy $E\left(\theta, \phi \right)$ is 
\begin{equation}
E(\theta,\phi)=E(s\bold{\hat{n}})= K_z s^2\cos^2\theta + K_ys^2\sin^2\theta\sin^2\phi,
\label{7.2}
\end{equation}
where $(\theta,\thinspace \phi)$ are the spherical coordinates of the magnetization direction $\bold{\hat{n}}$  and $K_z > K_y > 0$ are the anisotropy constants, $s$ is the particle's total spin, $M_0 =\gamma s$ is its magnetic moment, and $\gamma$ is related to the $g$ factor. The Euclidean action \eqref{7.1b} is similar to that in \eqref{2.7} except for an additional total derivative term. This term can be integrated out as:
\begin{equation}
 \int_{-\beta/2}^{\beta/2}d\tau\thinspace is\dot{\phi} = is\left[\phi(\beta/2)-\phi(-\beta/2) + 2n\pi\right],
\label{7.3}
\end{equation}
where $n$ is the winding number which counts the number of times the paths wrap around the north pole. As a total derivative, it has no contribution to the classical equations of motion, which can be derived by extremizing  the action with respect to $\theta(\tau)$ and $\phi(\tau)$. However, in computing the quantum transition amplitude, this term has a crucial property which makes the tunneling behavior of integral (bosons) and half-integral (fermions) spins to be different.

To compute the classical action for the instanton, we notice that the conservation of energy $E=0$ makes one of the trajectories imaginary, i.e $\cos\bar{\theta}$ or $\bar{\phi}$. Therefore only the first term in \eqref{7.1b} contributes to the instanton action. If $\cos\bar{\theta}$ is imaginary, we will obtain the real instanton trajectory in $\bar{\phi}$, thus, the instanton action will have two terms: an imaginary term plus an additional real term. The imaginary term of the instanton action is responsible for the suppression of tunneling while the real term gives the action for the instanton trajectory. On the other hand, if $\bar{\phi}$ is imaginary, we will obtain the real instanton trajectory in $\bar{\theta}$. Thus, the total action for the instanton becomes real and no suppression of tunneling can be found. 

Now, following the same approach in the previous chapter we see that $\cos\bar{\theta}$ is imaginary and hence the instanton action is 
\begin{equation}
S_{c} =  is\int_{0}^{\pi}d \bar{\phi} + B,
\label{8.0}
\end{equation}
where $B$ is given by 
\begin{equation}
B=s\sqrt{\lambda} \int_{0}^{\pi}d\bar{\phi}\thinspace\frac{\sin\bar{\phi}}{\sqrt{1-\lambda\sin^2\bar{\phi}}}= \ln\left(\frac{1+\sqrt{\lambda}}{1-\sqrt{\lambda}}\right)^{s}.
\label{8.1}
\end{equation}
Unlike the classical action found in the previous chapter, there is an additional imaginary contribution in \eqref{8.0} which comes from the total derivative. Now, consider for example the path $(\bar{\theta}(\tau), \thinspace \bar{\phi}(\tau))$ connecting the two anisotropy minima at $\bar{\phi}=0$ and $\bar{\phi}=\pi$, then owing to the symmetry of the action $S_0$ (that is excluding the total derivative term), the path $(\pi-\bar{\theta}(\tau), \thinspace -\bar{\phi}(\tau))$ will also solve the classical equations of motion and $B$ will be the same for both paths but the total derivative term will be reversed:  $ is\int_{-\infty}^{\infty}d\tau\thinspace\dot{\bar{\phi}} = is\int_{0}^{\pm\pi}d \bar{\phi}  =\pm is\pi$. In the semiclassical (small $\hbar$) approximation \cite{I, G}, the contributions of these two paths can be combined to give
\begin{equation}
e^{i\pi s}e^{-B/\hbar} + e^{-i\pi s}e^{-B/\hbar}=2\cos(\pi s)e^{-B/\hbar}.
\label{8.2}
\end{equation}
More appropriately, the tunneling rate can be obtained by summing over paths comprising of a sequence of instantons and anti-instantons winding over the barrier \cite{I, A, F},
\begin{equation}
\braket{\pi|e^{-\beta \hat{H}/\hbar}|0}  \propto e^{-\beta E_0}\sum_{m,n\geqslant 0}^{m+n \thinspace \text{odd}}\frac{(K\beta)^{m+n}}{m!n!}e^{is\pi(m-n)}e^{-B(m+n)/\hbar}=e^{-\beta E_0}\sinh\left[2K\beta\cos(\pi s)e^{-B/\hbar}\right]
\label{8.3},
\end{equation}
where $K$ is the fluctuation determinant \cite{G}, $m$ and $n$ are the number of instantons and anti-instantons in the paths respectively, $E_0=\hbar\omega/2$ is the zero-point energy in one well and $B$ is the action for the instanton. We can read off the tunneling rate (energy splitting) $\bigtriangleup E$ from \eqref{8.3}:
\begin{equation}
\Delta E = 4K\lvert\cos(\pi s)\rvert e^{-B/\hbar},
\label{8.4}
\end{equation}
where $B$ is given by \eqref{8.1}.

The $\cos(\pi s)$  is responsible for interference effect between instantons and anti-instantons. For integer spin $s$ (bosons), the interference is constructive $\cos(\pi s)= (-1)^s$, and the tunneling rate is non-zero, however, for half-odd-integer spin $s$ (fermions), the interference is destructive $\cos(\pi s)= 0$ and the tunneling rate vanishes.  As we pointed out above, the suppression of tunneling for half-odd-integer spins in this model is related to Kramers theorem. In subsequent examples, we shall break the time-reversal symmetry of the Hamiltonian by adding a magnetic field and show that the effect of suppression of tunneling still persists which is no longer related to Kramers theorem since the magnetic field breaks the time-reversal symmetry of the problem.

Another example is  that of a biaxial ferromagnetic particle with a magnetic field applied along the hard axis ($z$-direction) considered in \cite{K}. The quantum spin Hamiltonian is of the form
\begin{equation}
\hat{H} = K_z \hat{S_z^2} + K_y\hat{S_y^2} -\gamma H \hat{S_z} + \frac{\gamma^2 H^2}{4K_z},
\label{8.6}
\end{equation}
where $K_z > K_y >0$ are the anisotropy constants, $\gamma = g\mu_B  >0$, $H$ is the magnitude of applied field and $g$ is the spin $g$-factor. This Hamiltonian is no longer time reversal invariant due the presence of the magnetic field, so Kramers theorem is no longer applicable. We want to show that the suppression of tunneling still persists at certain values of the field. The classical anisotropy energy corresponding to this Hamiltonian is given by
\begin{equation}
E(\theta,\phi)=E(s\bold{\hat{n}}) = \braket{\bold{\hat{n}}|\hat{H}|\bold{\hat{n}}}= K_z s^2\cos^2\theta + K_ys^2\sin^2\theta\sin^2\phi -\gamma Hs\cos\theta +\frac{\gamma^2 H^2}{4K_z}.
\label{8.7}
\end{equation}
The minimum energy at $\phi=0,\thinspace \pi$ is obtained from the following conditions
\begin{equation}
\frac{d E}{d\theta}=0\quad \text{and}\quad\frac{d^2E}{d\theta^2}>0.
\label{8.7a}
\end{equation}
The first condition gives $u_0=\cos\theta_0 = H/H_c$ and the second condition is satisfied if  $H<H_c = 2K_z s/\gamma$ which is the case we are interested in. The classical anisotropy energy can thus be written as
\begin{equation}
E(\theta,\phi)=E(s\bold{\hat{n}}) = \braket{\bold{\hat{n}}|\hat{H}|\bold{\hat{n}}}= K_z s^2(\cos\theta -u_0)^2 + K_ys^2\sin^2\theta\sin^2\phi,
\label{8.7}
\end{equation}
which corresponds to two classical degenerate minima located at $\theta =\theta_0,\thinspace \phi= 0$ and $\theta =\theta_0,\thinspace \phi =\pi$. In order to compute the tunneling amplitude between these minima, we will follow the same approach as in model I.  The classical equations of motion are the same as \eqref{2.8} and \eqref{2.9} and the conservation of energy \eqref{2.13} still holds for this model with $E$ given by \eqref{8.7}.  Using the conservation of energy \eqref{2.13} we obtain the expression for $\cos\bar{\theta}$ in terms of $\bar{\phi}$
\begin{equation}
\cos\bar{\theta} = \frac{u_0 + i \lambda^{1/2}\sin\bar{\phi}(1-u_0^2 -\lambda\sin^2\bar{\phi})^{1/2}}{1-\lambda\sin^2\bar{\phi}},
\label{8.8}
\end{equation}
where $\lambda = K_y/K_z$. We have chosen the positive solution in \eqref{8.8} for convenience. Using this equation  and \eqref{8.7}, we can now eliminate $\bar{\theta}$ from \eqref{2.9} and obtain
\begin{equation}
\dot{\bar{\phi}}^2 =\omega_H^2\sin^2 \bar{\phi}(1-\lambda_H\sin^2 \bar{\phi}),
\label{8.9}
\end{equation}
where $\omega_H = 2s\sqrt{K_yK_z(1-u_0^2)}$ and $\lambda_H =  \lambda/(1-u_0^2)$. Upon integration we obtain the instanton solution
\begin{equation}
\bar{\phi}(\tau) =\pm\arccos\left(\frac{(\sqrt{1-\lambda_H})\tanh(\omega_H\tau)}{\sqrt{1-\lambda_H\tanh^2(\omega_H\tau)}}\right).
\label{9}
\end{equation}
Now, in this case the conservation of energy gives $\cos\bar{\theta}$ a real and imaginary terms. Thus, the classical action for this instanton path is:
\begin{equation}
S_{c} = i\pi\alpha +B,
\label{9.1}
\end{equation}
where
\begin{equation}
\alpha =\frac{s}{\pi} \int_{0}^{\pi}d\bar{\phi} \thinspace\left(1-\frac{u_0}{1-\lambda\sin^2\bar{\phi}}\right),
\label{9.2a}
\end{equation}
\begin{equation}
B =s\sqrt{\lambda}\int_{0}^{\pi}d\bar{\phi} \thinspace\frac{\sin\bar{\phi}\left(1-u_0^2-\lambda\sin^2\bar{\phi}\right)^{1/2}}{1-\lambda\sin^2\bar{\phi}}.
\label{9.2b}
\end{equation}
Following the same argument of pairing paths of opposite winding and summing over instantons and anti-instantons configurations, we obtain the transition amplitude
\begin{equation}
\braket{\pi|e^{-\beta \hat{H}/\hbar}|0}  \propto e^{-\beta E_0}\sum_{m,n\geqslant 0}^{m+n, \thinspace \text{odd}}\frac{(K\beta)^{m+n}}{m!n!}e^{i\pi\alpha(m-n)}e^{-B(m+n)/\hbar}=e^{-\beta E_0}\sinh\left[2K\beta\cos(\pi\alpha)e^{-B/\hbar}\right]
\label{9.3},
\end{equation}
and the tunneling rate  is then given by
\begin{equation}
\Delta E = 4K\lvert\cos(\pi\alpha)\rvert e^{-B/\hbar},
\label{9.4}
\end{equation}
where $\alpha$  and $B$ are easily obtained from \eqref{9.2a} and \eqref{9.2b} respectively:
\begin{align}
\alpha &=s\left( 1-\frac{u_0}{\sqrt{1-\lambda}}\right), \\
B&=2s\sqrt{\frac{u_0^2}{1-\lambda}}\arctanh\left(\sqrt{\frac{u_0^2\lambda}{(1-u_0^2)(1-\lambda)}}\right) + \ln \left(\frac{\sqrt{1-u_0^2}+\sqrt{\lambda}}{\sqrt{1-u_0^2}-\sqrt{\lambda}}\right)^s.
\label{9.5}
\end{align}
 The tunneling rate is thus suppressed whenever \cite{K}
 \begin{equation}
 u_0 =H/H_c = \sqrt{1-\lambda}\left(s-n-1/2\right)\slash s,
 \label{9.5}
 \end{equation}
 where $n$ is an integer.
\section{Suppression of tunneling in antiferromagnetic particles}
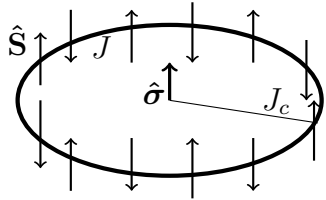
\begin{figure}[h!]
\centering
\begin{tikzpicture}
\draw[ultra thick](0,0) ellipse (2cm and 1cm);
\draw[->,very thick](0,0)--(0,0.5);
\draw[<-,thick](1.8,0)--(1.8,0.8);
\draw[<-,thick](1.9,0)--(1.9,-0.8);
\draw[->,thick](1.1,-0.5)--(1.1,-1.2);
\draw[<-,thick](0.3,-0.5)--(0.3,-1.3);
\draw[->,thick](-0.5,-0.5)--(-0.5,-1.2);
\draw(0,0)--(1.95,-0.3);
\draw(1.4,-0.01) node[]{$J_c$};
\draw(-0.2,0.1) node[]{$\boldsymbol{\hat{\sigma}}$};
\draw(-2.001,0.8) node[]{$\bold{\hat{S}}$};
\draw[<-,thick](-1.3,-0.5)--(-1.3,-1.2);
\draw[->,thick](-1.7,0)--(-1.7,-0.9);
\draw[->,thick](-1.7,0.2)--(-1.7,0.9);
\draw(-0.9,0.7) node[]{$J$};
\draw[<-,thick](-1.3,0.5)--(-1.3,1.2);
\draw[->,thick](-0.5,0.5)--(-0.5,1.2);
\draw[<-,thick](0.3,0.5)--(0.3,1.3);
\draw[->,thick](1.1,0.5)--(1.1,1.2);
\end{tikzpicture}
\caption{Antiferromagnetic ring coupled to an excess spin.}
\label{fig3.4}
\end{figure}

In this section we shall consider antiferromagnetic particles and investigate the effect of quantum phase interference. It is well known that tunneling rate in antiferromagnetic particles is much higher than that in ferromagnetic particles.  
We will begin with the analysis of a 1D antiferromagnetic ring with $N$ spins $\hat{\bold{S}}_j$ coupled to a central excess spin $\hat{\boldsymbol{\sigma}}$ with constant $J_c^{(j)}\equiv (-1)^jJ_c$ and a periodic boundary condition fig.\eqref{fig3.4}.
The model is considered in \cite{A}, the Hamiltonian of this system is
 \begin{equation}
\hat{H} = \sum_{j=1}^{N} \left[J\hat{\bold{S}}_{j}\cdot\hat{\bold{S}}_{j+1} +  K_z\hat{S}_{j}^{z2}+K_y\hat{S}_{j}^{y2} + J_c^{j}\hat{\bold{S}}_{j}\cdot\hat{\boldsymbol{\sigma}}\right],
\label{a}
\end{equation}
where $N $ is even.

Using the spin coherent state path integral formalism similar to that of
nonlinear sigma model \cite{J,F}, we have
\begin{equation}
 \braket{\bold{\hat{n}}_b|e^{- \beta\hat{H}/\hbar }|\bold{\hat{n}}_a}= \int   \mathcal{D}\left[\cos\theta \right]
 \mathcal{D}\left[\phi \right] e^{-S_E/\hbar}.
\label{b}
\end{equation}

The effective Euclidean action is of the form
\begin{equation}
S_E = \int_{0}^{\beta}d\tau \left(\frac{\chi_\perp}{8\mu_B ^2}\left(\dot{\theta}_N^2 + \dot{\phi}_N^2\sin^2\theta_N\right) + E(\theta_N,\phi_N)+ sNJ_c\bold{\hat{N}}\cdot\bold{\hat{n}_\sigma}+ i\sigma\dot{\phi}_\sigma(1-\cos\theta_\sigma)\right),
\label{c}
\end{equation}
where the Ne\`el vector $\bold{\hat{N}}$ and the unit vector $\bold{\hat{n}_\sigma}$ are expressed in spherical coordinates.
The classical anisotropy energy $E(\theta,\phi)$ is given by:
\begin{equation}
E(\theta_N,\phi_N)=\tilde{ K_z} s^2\cos^2\theta_N + \tilde{K_y}s^2\sin^2\theta_N\sin^2\phi_N,
\label{d}
\end{equation}
where $\tilde{K}_{y,z} = NK_{y,z}$. and $\chi_\perp=N\mu_B ^2/J$. For $\hat{\boldsymbol{\sigma}} =0$, the last two terms in \eqref{c} vanish, thus, the action has only a real part and no suppression of tunneling is expected. However, for $\hat{\boldsymbol{\sigma}}\neq0$, the action has both real and imaginary parts and one should expect suppression of tunneling. In this case, we will simplify the problem by assuming $\bold{\hat{N}}\perp\bold{\hat{n}_\sigma}$, $\tilde{K_z}>>\tilde{K_y}$ and setting $\theta_N=\theta_\sigma =\theta$, $\phi_N=\phi_\sigma=\phi$. In the limit of this strong transverse anisotropy, $\theta$ does not fluctuate very far away from $\pi/2$, then we can write $\theta = \pi/2 - \vartheta$ and expand the effective action to second order in $\vartheta$, we obtain from \eqref{d} and \eqref{c}
\begin{equation}
S_E = \int_{0}^{\beta}d\tau \left(\frac{\chi_\perp}{8\mu_B ^2} \dot{\phi}^2 + \vartheta \mathcal{G}^{-1}\left[\phi\right]\vartheta + \tilde{K_y}\sin^2\phi + i\sigma\dot{\phi}(1-\vartheta)\right),
\label{e}
\end{equation}
where $\mathcal{G}^{-1}\left[\phi\right]= \left(\tilde{K_z}-\tilde{K_y}\sin^2\phi +\frac{\chi_\perp}{8\mu_B ^2}(\partial_\tau ^2-\dot{\phi}^2)\right)\approx \tilde{K_z}$ and $\mathcal{D}\left[\cos\theta \right]\approx \mathcal{D}\left[\vartheta \right]$. Integrating out $\vartheta$ in \eqref{b} we obtain
\begin{equation}
 \braket{\bold{\hat{n}}_b|e^{- \beta\hat{H}/\hbar }|\bold{\hat{n}_a}}= \int    
 \mathcal{D}\left[\phi \right] e^{-S_E^{eff}/\hbar},
\label{f}
\end{equation}
where
\begin{equation}
S_E^{eff} = \int_{0}^{\beta}d\tau\left(\frac{I}{2}\dot{\phi}^2 + \tilde{K_y}\sin^2\phi + i\sigma\dot{\phi} \right),
\label{g}
\end{equation}
and $I=   \chi_\perp/4\mu_B ^2 +\sigma^2/2\tilde{K_z}
 $.
 The first integral of the classical equation  of motion gives
\begin{equation}
\frac{I}{2}\dot{\bar{\phi}}^2 - V(\bar{\phi})  =0,
\label{h}
\end{equation}
where $V(\bar{\phi})= \tilde{K_y}\sin^2\phi$.

The instanton solution of \eqref{h} corresponds to the tunneling of the Ne\`el vector through a potential barrier from $\phi=0$ at $\tau =-\infty$ to $\phi = \pi$ at $\tau=\infty$ along clockwise and anticlockwise paths. The solution is
\begin{equation}
\bar{\phi}(\tau) = \pm 2\arctan(e^{\omega_0 \tau}),
\label{i}
\end{equation}
where $\omega_0 = \sqrt{2\tilde{K_y}/I}$.
In order to obtain the tunneling splitting we follow the usual procedure of  summing over instantons and anti-instantons configurations, this gives the tunneling rate (energy splitting)  
\begin{equation}
\Delta E = 4K\lvert\cos(\pi\sigma)\rvert e^{-B/\hbar},
\label{k}
\end{equation}
$B$ can be obtained from \eqref{g}:
\begin{equation}
B=2I\omega_0.
\label{l}
\end{equation}
Therefore, we see that quenching of tunneling rate for half-odd-integer spins persists in antiferromagnetic particle.

The final model we will look at is the  antiferromagnetic particles describe by the Ne\`el vector of two collinear sublattices whose spins are coupled by strong interaction. In the absence of the magnetic field, the two spins are opposite to each other $S_1=-S_2$, so the total spin vanishes. We will consider a biaxial antiferromagnetic particle of two collinear ferromagnetic sublattices with a small non-compensation and assume that it possesses an $x$-easy-axis and $xy$ easy plane and a magnetic field $h$ is applied along the hard axis ($z$-axis). The Hamiltonian operator for this model is given by \cite{B}
\begin{equation}
\hat{H} = J\hat{\bold{S}}_1\cdot\hat{\bold{S}}_2 +\sum_{\alpha=1,2}\left(K_z\hat{S}_{\alpha}^{z2}+K_y\hat{S}_{\alpha}^{y2}-\gamma h\hat{S}_{\alpha}^{z}\right),
\label{10.0}
\end{equation}
where $K_z > K_y >0$ are the anisotropy constants, $J$ is the exchange constant, $\gamma = g\mu_B  >0$, $h$ is the magnitude of applied field and $g$ is the spin $g$ factor.
The spin operators in the two sublattices $\hat{\bold{S}}_1$ and $\hat{\bold{S}}_2$ obey the usual commutator relation
\begin{equation}
 [\hat{S}_{\alpha}^i, \hat{S}_{\beta}^j ]=i\epsilon_{ijk}\delta_{\alpha\beta}\hat{S}_{\gamma}^k,
\label{10.1}
\end{equation}
where $i,j,k = x,y,z$ and $\alpha, \beta =1,2$. Using the spin coherent state path integral representation we have
\begin{equation}
 \braket{\bold{\hat{n}}_b|e^{- \beta \hat{H}/\hbar}|\bold{\hat{n}}_a}= \int  \prod_{\alpha=1,2}\mathcal{D}\left[\cos\theta_{\alpha}\right]
 \mathcal{D}\left[\phi_{\alpha}\right] e^{-S_E/\hbar},
\label{10.2}
\end{equation}
where $S_E = \int_{-\beta/2}^{\beta/2}d\tau\mathcal{L}_E$ and
\begin{equation}
\begin{split}
\mathcal{L}_E &= \sum_{\alpha=1,2}\left[is_{\alpha}\dot{\phi}_{\alpha}(1-\cos\theta_\alpha)+K_zs_\alpha ^2\cos^2\theta_\alpha +K_ys_\alpha ^2 \sin^2\theta_\alpha\sin^2\phi_\alpha -\gamma hs_\alpha\cos\theta_\alpha\right]\\&
+Js_1s_2\left[\sin\theta_1\sin\theta_2\cos(\phi_1-\phi_2)+\cos\theta_1\cos\theta_2\right].
\label{10.3}
\end{split}
\end{equation}
Since we are looking for quantum transitions between macroscopic states, only low-energy trajectories with almost antiparallel $s_1$ and $s_2$ contribute to the path integral, therefore we can replace $\theta_2$ and $\phi_2$ by $\pi-\theta_1-\epsilon_\theta$ and $\pi+\phi_1+\epsilon_\phi$, where $\epsilon_\theta,\epsilon_\phi << 1$ denotes small fluctuations. Plugging this into \eqref{10.3} and expanding to second order in $\epsilon$ we obtain
\begin{equation}
\begin{split}
\mathcal{L}_E &=  is_0\dot{\phi}-i\tilde{s}\dot{\phi}\cos\theta + 2(K_zs^2\cos^2\theta +K_ys^2\sin^2\theta\sin^2\phi)-\gamma h\tilde{s}\cos\theta\\&+ \left[-is\dot{\phi}\sin\theta -K_zs^2\sin2\theta + K_ys^2\sin^2\phi \sin2\theta-\gamma hs\sin\theta\right]\epsilon_\theta\\&+ (K_s s^2\sin^2\theta\sin2\phi)\epsilon_\phi + is\left[(1+\cos\theta)-\sin\theta\epsilon_\theta\right]\dot{\epsilon_\phi}\\&
+ (E_{\theta\theta}\epsilon_{\theta}^2 +E_{\theta\phi}\epsilon_{\theta}  \epsilon_{\phi} + E_{\phi\phi}\epsilon_{\phi}^2 )
\label{10.4}
\end{split}
\end{equation}
where
\begin{equation}
\begin{split}
 E_{\theta\theta}&=\frac{Js^2}{2} -\frac{is}{2}\dot{\phi}\cos\theta +K_ys^2\cos2\theta\sin^2\phi -\gamma \frac{hs}{2}\cos\theta\\
 E_{\phi\phi} &= \frac{Js^2}{2}\sin^2\theta +K_ys^2\sin^2\theta\cos2\phi\\
 E_{\theta\phi}&=K_ys^2\sin2\theta\sin2\phi
\label{10.5}
\end{split}
\end{equation}
and $\tilde{s}=s_1-s_2$, $s_0 = s_1 +s_2=2s$, we have set $s_1=s_2=s$ except in the terms containing $s_1-s_2$, $(\theta_1,\phi_1)=(\theta,\phi)$. Working out the Gaussian integration over $\epsilon_\theta$ and $\epsilon_\phi$, Eq.\eqref{10.2} reduces to
\begin{equation}
 \braket{\bold{\hat{n}}_b|e^{- \beta \hat{H}/\hbar }|\bold{\hat{n}}_a}= \int   \mathcal{D}\left[\cos\theta\right]
 \mathcal{D}\left[\phi \right] \exp\left(-\int d\tau\mathcal{L}_E^{eff} \right),
\label{10.6}
\end{equation}
where
\begin{equation}
\begin{split}
\mathcal{L}_E ^{eff} &=  i\frac{\tilde{m}_0}{\gamma}\dot{\phi}-i\frac{\tilde{m}}{\gamma}\dot{\phi}\cos\theta -i\frac{\chi_{\perp}}{\gamma}h\dot{\phi}\sin^2\theta + \frac{\chi_{\perp}}{2\gamma^2}(\dot{\theta}^2+\dot{\phi}^2\sin^2\theta)\\&+ \tilde{K_z}\cos^2\theta +\tilde{K_y}\sin^2\theta\sin^2\phi- h\tilde{m}\cos\theta,
\label{10.7}
\end{split}
\end{equation}
and $\tilde{m}_0=\gamma s_0, \quad \tilde{m}=\gamma(s_1-s_2),\quad \chi_\perp= \gamma^2/J,\quad \tilde{K_z}= 2K_zs^2,\quad \tilde{K_y}=2K_ys^2$.
We can find an approximate solution of the tunneling rate if we make the assumption that $\tilde{K_z}>>\tilde{K_y}$ and therefore $\theta$ does not fluctuate very far away from $\pi/2$, thus we can do the expansion $\theta=\pi/2 -\vartheta$, then the effective Lagrangian to second order in $\vartheta$ becomes
\begin{equation}
\begin{split}
\mathcal{L}_E ^{eff} &=  i\frac{\tilde{m}_0-\chi_{\perp}h}{\gamma}\dot{\phi} + \frac{\chi_{\perp}}{2\gamma^2}\dot{\phi}^2 +\tilde{K_y}\sin^2\phi -\frac{\chi_\perp}{2}h^2 +\vartheta\mathcal{G}^{-1}\left[\phi\right]\vartheta - (\frac{i\tilde{m}}{\gamma}\dot{\phi} + \tilde{m}h)\vartheta, 
\label{10.8}
\end{split}
\end{equation}
where $\mathcal{G}^{-1}\left[\phi\right] = \tilde{K_z}-\tilde{K_y}\sin^2\phi -\frac{\chi_\perp}{2\gamma^2}(\partial_\tau +\dot{\phi}^2)+i\frac{\chi_\perp}{\gamma}h\dot{\phi}+\frac{\chi_\perp}{2}h^2 \approx \tilde{K_z}$. Performing the Gaussian integration over $\vartheta$ we have
\begin{equation}
 \braket{\bold{\hat{n}}_b|e^{- \beta\hat{H}/\hbar }|\bold{\hat{n}}_a}= \int    
 \mathcal{D}\left[\phi \right]e^{-S_E^{eff}/\hbar},
\label{10.9}
\end{equation}
where
\begin{equation}
\begin{split}
S_E^{eff} &= i\Theta\int d\tau \dot{\phi} +\int d\tau\left( \frac{I}{2}\dot{\phi}^2 + V(\phi)\right).
\label{11}
\end{split}
\end{equation}
The constants are given by: $I = I_a +I_f$, where $I_a = \tilde{m}^2/(2\gamma^2\tilde{K_z})$ and $I_f=\chi_\perp/\gamma^2$ are the effective antiferromagntic and ferromagnetic moments of inertia, $\Theta = m_0 -I\gamma h$. $V(\phi)= \tilde{K_y}\sin^2\phi$.
The first integral of the classical equation of motion is similar to that of the previous model. It is given by
\begin{equation}
\frac{I}{2}\dot{\bar{\phi}}^2 - V(\bar{\phi})  =0,
\label{11.1}
\end{equation}
with the instanton solution
\begin{equation}
\bar{\phi}(\tau) = \pm 2\arctan(e^{\omega_0 \tau}),
\label{11.2}
\end{equation}
where $\omega_0 = \sqrt{2\tilde{K_y}/I}$.
This solution \eqref{11.2} corresponds to the tunneling of the Ne\`el vector through a potential barrier from $\phi=0$ at $\tau =-\infty$ to $\phi = \pi$ at $\tau=\infty$ along clockwise and anticlockwise paths.

The tunneling rate follows the usual procedure, in this case we have
\begin{equation}
\Delta E = 4K\lvert\cos(\pi\Theta)\rvert e^{-B/\hbar},
\label{11.4}
\end{equation}
where $\Theta =m_0 -I\gamma h$ and $B$ can be obtained from \eqref{11}:
\begin{equation}
B=2I\omega_0.
\label{11.5}
\end{equation}
The tunneling is suppressed whenever $h = (m_0-n-1/2)/I\gamma$, where $n$ is an integer \cite{O}. If $h=0$, we see that the tunneling splitting is suppressed for half-odd-integer $m_0$ but survives for integer $m_0$ \cite{B}. Therefore suppression of tunneling due to quantum phase interference occurs 
in both ferromagnetic and antiferromagnetic spin particles.

\chapter{Summary}
\section{Conclusion}
In conclusion, we have presented a comprehensive study of macroscopic quantum coherence and tunneling of spins in ferromagnetic and antiferromagnetic spin systems with arbitrary magnetic anisotropy energy using the path integral method. We computed explicitly the actions for the instanton and the bounce paths and the corresponding tunneling amplitudes in these systems. We obtained the crossover temperatures at which quantum transition is dominated by thermal hopping over the energy barrier. In the presence of a magnetic field, we found that the crossover temperature depends on the anisotropy field. Thus, for a particle with $H_c>>1$Tesla, a reasonably low temperature is required. It is experimentally possible to determine the crossover temperature due to its weak dependence on $\epsilon$. However, the action for this systems was insufficient to explain the spin-parity effect in spin systems. 

Using the spin coherent state path integral, we obtained an additional contribution to the action leading to a topological phase factor. We showed that this topological phase in the spin tunneling amplitude leads to both destructive and constructive interference between tunneling paths. In the case of destructive interference, the tunneling rate (energy splitting) is zero which leads to the suppression of tunneling for half-odd-integer spins while for constructive interference, the tunneling rate is non-zero for integer spins. We showed that the quenching of the tunneling rate for half-odd-integer spins is related to Kramers theorem if the Hamiltonian is time-reversal invariant, however quenching of tunneling rate persists when the Hamiltonian is not time-reversal invariant, in this case Kramers degeneracy is removed and the quenching of the tunneling rate is no longer related to Kramers theorem. We also showed that this spin parity effect occurs both in ferromagnetic and antiferromagnetic particles. Moreover, suppression of tunneling also occurs whenever the Hamiltonian commutes with its spin variables. Most of the results obtained in this essay have been verified experimentally and the research on macroscopic quantum tunneling of spins is on-going. 
 
\renewcommand{\bibname}{References}
\nocite{*}
\bibliographystyle{amsplain}
\bibliography{Pre-doctoral_exam_2012}

\providecommand{\bysame}{\leavevmode\hbox to3em{\hrulefill}\thinspace}
\providecommand{\MR}{\relax\ifhmode\unskip\space\fi MR }
\providecommand{\MRhref}[2]{%
  \href{http://www.ams.org/mathscinet-getitem?mr=#1}{#2}
}
\providecommand{\href}[2]{#2}
\begin{thebibliography}{10}

\bibitem{R}
E.~M. Chudnovsky and B.~Barbara, Physics Letters A. \textbf{145} (1990), 205.

\bibitem{E}
E.~M. Chudnovsky, B.~Barbara, and P.~C~.E Stamp, International Journal of
  Modern Physics B. \textbf{6} (1992), 1355.

\bibitem{D}
E.~M. Chudnovsky and L.~Gunther, Phys. Rev. Lett. \textbf{60} (1988), 661.

\bibitem{C}
E.~M. Chudnovsky, O.~Iglesias, and P.~C~.E Stamp, Phys. Rev. B \textbf{46}
  (1992), 5392.

\bibitem{B}
Eugene.~M. Chudnovsky, Journal of Magnetism and Magnetic Material \textbf{140}
  (1995), 1821.

\bibitem{I}
Sidney Coleman, \emph{Aspects of symmetry}, Cambeidge University Press, {1985}.

\bibitem{G}
Jr. Curtis G.~Callan and Sidney Coleman, Phys. Rev. D \textbf{16} (1977), 1762.

\bibitem{A}
Jan~Von Delft and Christopher~L. Henley, Phys. Rev. Lett. \textbf{69} (1992),
  3236.

\bibitem{L}
R.~P Feynman and A.~R Hibbs, \emph{Quantum mechanics and path integrals},
  McGraw-Hills, New York, {1965}.

\bibitem{J}
Eduardo Fradkin, \emph{Field theories of condensed matter systems},
  Addison-Wesley, {1991}.

\bibitem{K}
A.~Garg, EuroPhys. Lett. \textbf{22} (1993), 205.

\bibitem{H}
John~R. Klauder, Phys. Rev. D \textbf{19} (1978), 2349.

\bibitem{F}
Daniel Loss, David~P. DiVincenzo, and G.~Grinstein, Phys. Rev. Lett.
  \textbf{69} (1992), 3232.

\bibitem{O}
Yi-Hang Nie, Yan-Hong Jin, J-Q Liang, and H~J~W Muller-Kirsten, J. Phys.
  Condens. Matter \textbf{12} (2000), L87.

\bibitem{T}
J.~M. Radcliffe, J. Phys. A \textbf{4} (1970), 313.

\bibitem{N}
Subir Sachdev, \emph{Quantum phase transitions 2nd edition}, Cambridge
  University Press, {2011}.

\bibitem{M}
Wei-Min Zhang, Da~Hsuan Feng, and Robert Gilmore, Rev. Mod. Phys \textbf{62}
  (1990), 867.

\end{thebibliography}
\addcontentsline{toc}{chapter}{References}
\end{document}